\documentclass[acmtog]{acmart}
\acmSubmissionID{1376}

\usepackage{booktabs} 

\renewcommand\footnotetextcopyrightpermission[1]{} 

\usepackage{pifont}
\usepackage{graphicx}
\usepackage{subcaption}
\usepackage{multirow}
\usepackage{makecell, multicol, diagbox}
\usepackage{xcolor, colortbl}
\usepackage{array}
\newcolumntype{C}[1]{>{\centering\arraybackslash}p{#1}}
\newcommand{\winner}[1]{\cellcolor{red!20}{#1}}
\newcommand{\runner}[1]{\cellcolor{orange!20}{#1}}
\newcommand{\thirdp}[1]{\cellcolor{yellow!20}{#1}}
\newcommand{\normal}[1]{#1}

\def\figurePath{fig/}
\def\myfigure#1#2#3{
\begin{figure}[htb]\centering\includegraphics[width = \linewidth]{\figurePath#2}\caption{#3}\label{fig:#1}
\end{figure}}

\def\mycfigure#1#2#3{
\begin{figure*}[htb]\centering\includegraphics*[clip, width = \linewidth]{\figurePath#2}\caption{#3}\label{fig:#1}
\end{figure*}}

\usepackage{gensymb}

\usepackage[ruled]{algorithm2e} 

\begin{document}
\title{Free Your Hands: Lightweight Turntable-Based Object Capture Pipeline }


\author{Jiahui Fan}
\orcid{0000-0003-0871-7615}
\affiliation{
    \institution{Nanjing University of Science and Technology}
    \country{China}
}
\email{fjh@njust.edu.cn}

\author{Fujun Luan}
\orcid{0000-0001-5926-6266}
\affiliation{
    \institution{Adobe Research}
    \country{USA}
}
\email{fluan@adobe.com}

\author{Jian Yang$^\dagger$}
\orcid{0000-0003-4800-832X}
\affiliation{
    \institution{Nanjing University of Science and Technology}
    \country{China}
}
\email{csjyang@njust.edu.cn}

\author{Milo\v{s} Ha\v{s}an}
\orcid{0000-0003-3808-6092}
\affiliation{
    \institution{Adobe Research}
    \country{USA}
}
\email{milos.hasan@gmail.com}

\author{Beibei Wang$^\dagger$}
\orcid{0000-0001-8943-8364}
\affiliation{
    \institution{Nanjing University}
    \country{China}
}
\email{beibei.wang@nju.edu.cn}

\renewcommand\shortauthors{Fan J. et al}

\begin{teaserfigure}
   \centering
    \includegraphics[width=\textwidth]{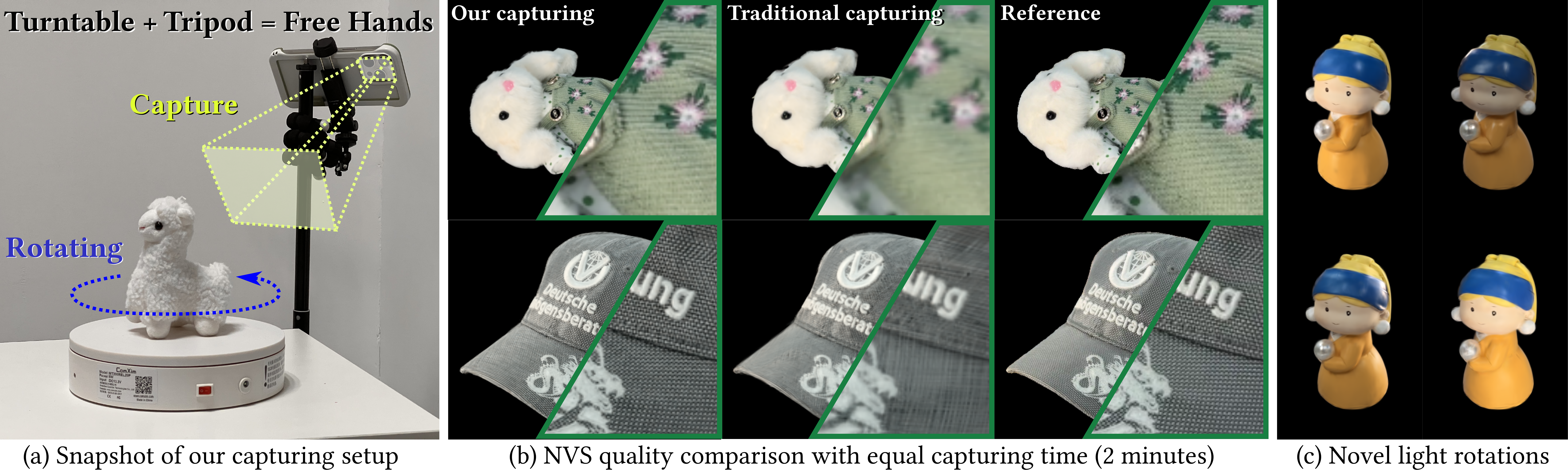}
    \caption{
    We present a lightweight object capture pipeline to reduce manual effort, while increasing reconstruction quality and additionally supporting novel light rotations. (a) We use a consumer turntable to carry the object and a tripod to hold the camera, automatically capturing dense samples from various views and lighting rotations, obtaining hundreds of high-quality captures within 2 minutes. (b) We develop a rotation-conditional neural radiance representation, achieving NVS results with less blur and clearer details. (c) With our new representation, we can synthesize results under novel views and lighting rotations.
    }
    \label{fig:teaser}
\end{teaserfigure}

\begin{abstract}
Novel view synthesis (NVS) from multiple captured photos of an object is a widely studied problem. Achieving high quality typically requires dense sampling of input views, which can lead to frustrating manual labor. Manually positioning cameras to maintain an optimal desired distribution can be difficult for humans, and if a good distribution is found, it is not easy to replicate. Additionally, the captured data can suffer from motion blur and defocus due to human error. In this paper, we use a lightweight object capture pipeline to reduce the manual workload and standardize the acquisition setup, with a consumer turntable to carry the object and a tripod to hold the camera. Of course, turntables and gantry systems have been frequently used to automatically capture dense samples under various views and lighting conditions; the key difference is that we use a turntable under \emph{natural environment lighting}. This way, we can easily capture hundreds of valid images in several minutes without hands-on effort. However, in the object reference frame, the light conditions vary (rotate); this does not match the assumptions of standard NVS methods like 3D Gaussian splatting (3DGS). We design a neural radiance representation conditioned on light rotations, which addresses this issue and allows rendering with novel light rotations as an additional benefit. We further study the behavior of rotations and find optimal capturing strategies. We demonstrate our pipeline using 3DGS as the underlying framework, achieving higher quality and showcasing the method's potential for novel lighting and harmonization tasks.


\end{abstract}

%
%

%
%

\keywords{Object capturing, Neural appearance, Gaussian splatting, }

\maketitle

\section{Introduction}

Creating realistic 3D assets from real-world objects is a long-standing challenge of computer graphics, with applications to e-commerce, entertainment, digital heritage, and more. A typical pipeline captures multiple views (typically hundreds) of an object under fixed lighting and transforms the captured images into a 3D asset representation, which allows novel view synthesis (NVS). Various methods based on Neural radiance fields (NeRF)~\cite{mildenhall2021nerf} and 3D Gaussian splatting (3DGS)~\cite{kerbl20233d} can be used for this purpose. Still, high-quality capture requires careful manual labor and time. With hand-held cameras, one can either capture discrete images or video sequences, while in both cases, the quality of captured data can suffer from issues like motion blur and defocus, especially for capturing videos, as shown in Fig.~\ref{fig:static}. Moreover, the constraints on a reasonable sampling of viewpoints may be obvious to researchers, but manually positioning cameras to maintain the desired distribution can be difficult for average users. 
    
Extensive research~\cite {yu2024gsdf, chen2022tensorf, barron2021mip, Attal_2023_CVPR} improves upon the above NVS pipelines. For example, various lighting setups can be employed, including unknown environment lighting~\cite{zhang2020nerf++, wang2021nerf}, or a point light that is collocated with the camera~\cite{bi20203dcapture, bi2020deep, bi2020neural}, or separate from the camera~\cite{gao2020deferred, bi2024rgs}. However, controlled lighting generally requires a dark room and more involved / expensive setups, and becomes out of reach for most non-expert users. Several works propose to utilize sparse views and cross-scene feed-forward inference techniques~\cite{MVSplat:2024:Chen,charatan2024pixelsplat}. However, the reconstructed quality necessarily degrades without dense samples. Dense images can be captured with various specialized devices \cite{ye2024pvp, Ma:2021:TRACE, kang2023differentiable}, which can reduce the required human labor and precision, but are not available off-the-shelf.

\myfigure{static}{static.pdf}{Illustration of some the common problems of ordinary hand-held captured pipeline. The final NVS quality can suffer from the imperfectly sampled distribution and blurry images.}


In this paper, we present \textit{Free Your Hands}: a repeatable and robust lightweight object capture pipeline designed to reduce manual workload and standardize the acquisition process, while increasing reconstruction quality and offering novel illumination rotations as an additional benefit. We use a consumer turntable to carry the object and a tripod to support the camera. As the turntable rotates, we automatically capture dense samples from various views under rotating environment illumination (as observed from the object's frame of reference). This defines a controllable and repeatable sampling trajectory, and the tripod-mounted camera minimizes motion blur and defocus in video sequences, compared to stationary objects plus hand-held cameras. With this setup, we can easily capture hundreds of valid images in 2 minutes without much manual effort, while typically under 50 valid images with hand-held multi-view acquisition can be captured in the same time. We calibrate camera poses with a standard pipeline~\cite{schonberger2016structure}.

The captured images could theoretically be used in existing NVS frameworks (NeRF or 3DGS), but the challenge lies in the varying (rotating) light conditions that break the assumptions of these techniques and are harmful to reconstruction quality. Therefore, we define a new radiance representation conditioned on light rotations. Furthermore, we investigate the relationship between the rotation of objects and the NVS problem, providing experiments and analysis to find an optimal capturing setup that balances human labor and NVS reconstruction quality.

Of course, a turntable is not novel in computer graphics and appearance capture. Previous works focus on calibration and pose estimation from such data \cite{pusztai2016turntable, elms2024event, cheng2023accidental}, or reconstruction of complex light transport with controlled light sources\cite{wu2018full}. Methods that use point light illumination in a dark room \cite{bi20203dcapture, bi2020deep, bi2020neural, zeng2023nrhints} could use a turntable with minimal changes: if camera and light poses with respect to the object are known, it does not matter whether the object is rotating. However, we are the first to investigate the relationship of a turntable under \emph{natural environment illumination} to the NVS problem, and to practically develop such a capture pipeline. 
To summarize, our main contributions include:
\begin{itemize}
    \item a novel lightweight object capture pipeline that uses a turn\-table under natural environment illumination, combined with a tripod to reduce manual workload and standardize the acquisition process,
    \item fitting a rotation-conditioned radiance field representation, allowing accurate NVS as well as novel light rotation for applications such as harmonization,
    \item two rotation-based sampling strategies tailored for our capture pipeline, and
    \item a dataset of objects with varying material properties captured and reconstructed under different rotating configurations with our pipeline for downstream research.
\end{itemize}

\section{Related work}


\paragraph{Novel view synthesis}

Novel view synthesis (NVS) aims to generate new images from view directions that were not originally observed. Extensive research on 3D representations has been proposed to enable realistic novel view rendering. 
Notably, NeRF \cite{mildenhall2021nerf}, 3DGS \cite{kerbl20233d} and their follow-ups have garnered significant attention due to their powerful representation capabilities.
NeRF models radiance using integrals over a ray passing through a volume, and addresses view-dependent radiance using multilayer perceptrons (MLPs) conditioned on the view ray direction; this is also true in most follow-up methods \cite{chen2022tensorf, mueller2022instant}. Enhanced anti-aliasing techniques~\cite{barron2021mip, barron2022mip, barron2023zip, zhang2020nerf++} and improved reflectance modeling \cite{verbin2022ref, Attal_2023_CVPR} have further refined the quality and performance of NeRF-based representations. 
3DGS \cite{kerbl20233d} employs anisotropic Gaussians to represent scenes, allowing for great adaptivity to actual geometric content and enabling real-time, highly detailed renderings. The view dependence of radiance is represented using spherical harmonics, which is even more limited than MLP-based view-dependence.

\paragraph{Camera pose calibration}
These above techniques rely on posed images to densely reconstruct 3D scenes, with the quality of the input images significantly influencing the final rendering results. Structure from Motion (SfM) methods \cite{hartley2003multiple, schonberger2016structure, mur2015orb, taketomi2017visual}, particularly COLMAP \cite{schonberger2016structure}, are commonly employed to calibrate camera poses and provide initialization for point-based techniques.
Although the quality of input data is typically assumed to be sufficient for COLMAP to succeed, this assumption is frequently invalid. Consequently, standardizing the data acquisition process and minimizing human error is beneficial to ensure robust and reliable performance in NVS applications.
Recently, VGGT~\cite{wang2025vggt} introduces an end-to-end transformer model to automatically predict all key 3D attributes of a scene by feeding several multi-view images, providing a powerful solution that is compatible with multiple tasks in 3D computer vision.

\section{Methods}
\label{sec:method}

In this section, we first briefly review the traditional multi-view NVS strategy with static objects (Sec.~\ref{sec:static_sampling}), and then introduce the insight of covering the "2D" sampling domain of views and light rotations with our turntable capturing pipeline (Sec.~\ref{sec:rotating_sampling}). Next, to achieve NVS in the proposed sampling domain, we present a conditional radiance representation (Sec.~\ref{sec:mlp}), and we further investigate the relationship between rotation and NVS quality to find an optimal setup (Sec.~\ref{sec:swing_sampling}).

\subsection{Static sampling}
\label{sec:static_sampling}

\myfigure{space1d}{space1d.pdf}{The traditional capture strategy visualized in a 2D sampling domain. The object is stationary, and the camera is placed at multiple positions for a dense sampling of the NVS problem.}

To create a 3D asset from a real-world object for novel view synthesis, a common approach is to capture hundreds of images by moving the camera around the object under static light conditions and then reconstruct the radiance field with NeRF or 3DGS or their variants. The entire pipeline can be formulated as follows:
\begin{align}
   L_{\mathrm{static}} &= \mathcal{G}(\mathbf{V} \vert I),\\
   \{ \mathbf{V} \}  &= \{ \mathbf{V}_i | i = 1, 2, ..., M \},
    \label{eq:typical}
\end{align}
where $L_{\mathrm{static}}$ represents the radiance field, $\mathcal{G}$ denotes a reconstruction operator, which could be based on NeRF or 3DGS or their variants, $\mathbf{V}$ is a set of camera views described by camera intrinsics and extrinsics, and $I$ is the illumination condition under which the images are captured. The camera is placed at $M$ different positions. The target of NVS is to predict the radiance $L^\prime_{\mathrm{static}}$ at a novel viewpoint $v^\prime$, while the lighting $I$ stays the same. We can  visualize this formulation in the object frame of reference as a column in a "2D" sampling domain, where the two "dimensions" represent the camera view and the light rotation angle, as shown in Fig.~\ref{fig:space1d}. Note that for simplicity, we visualize the camera views as one vertical dimension. Since the object (and therefore the light rotation) stays fixed throughout the capture process, we call it the \textit{static} sampling strategy.

To get a faithful description of the entire "column", it is necessary to densely sample the viewpoints. The user should ensure a reasonable distribution of sampled views and be careful not to introduce motion blur and defocus issues; this is a difficult task fo non-experts. The question is: can we free our hands while maintaining high-quality NVS reconstruction?

\subsection{Rotating sampling}
\label{sec:rotating_sampling}

Instead of moving the camera, we can rotate the object itself. An effective approach is to use a turntable to hold the object, allowing it to rotate automatically. We can position the camera on a tripod at a fixed location to ensure video (or image burst) sequences without blur. As the turntable rotates, high-quality sequences are captured and automatically ensure a uniform sampling along the trajectory. We call this the \textit{rotating} sampling strategy. This strategy can be illustrated in the 2D sampling domain as well, but as a diagonal line, as shown in Fig.~\ref{fig:space2d} (a). Note again, we visualize the camera views as one vertical dimension; the diagonal line represents the intuition that the camera view and light rotation, in the object frame of reference, are "tied" together.

\myfigure{space2d}{space2d.pdf}{Our \textit{rotating} sampling strategy illustrated in the 2D sampling domain. The object is rotating on a turntable, and the camera is placed at a fixed position (a) or multiple positions (b). Each camera position leads to an individual diagonal line in the sampling domain.}


Suppose we place the camera at $M$ different locations, and the turntable completes a full rotation each time. This strategy can be illustrated as multiple diagonal lines, as shown in Fig.~\ref{fig:space2d} (b). Here the intuition is that the camera view and light rotation, in the object frame of reference, are "tied" together differently for each turntable sequence.
From each video sequence corresponding to a camera position, we sample $N$ images, resulting in $MN$ images for reconstruction. Now the reconstruction formulation becomes:
\begin{align}
   L_{\mathrm{rotating}} & = \mathcal{G}(\mathbf{V}, \mathbf{I}), \\
   \{ \mathbf{V} \} & = \{ \mathbf{V}_i | i = 1, 2, ..., M \}, \\
   \{ \mathbf{I} \} & = \{ \mathbf I_j | j = 1, 2, ..., N \},
\end{align}
where $L_\mathrm{rotating}$ is a radiance field that is conditioned on view directions and light rotations. 



With $M$ different camera positions and a full 360-degree turntable rotation per camera position, we have $M$ diagonal lines in this 2D sampling domain. With well distributed camera positions, we can get good coverage of the 2D domain with multiple diagonal lines, which allows high-quality novel view and light rotation synthesis. (We can further optimize this strategy by letting the turntable rotate for a shorter angle per camera view, which we will present in Sec.~\ref{sec:swing_sampling}.)

\subsection{Conditional radiance field with rotating lighting}
\label{sec:mlp}

The next question is how to reconstruct the conditional radiance field from these captured images with light rotations. 
The original formulation for radiance field reconstruction, as presented in Eqn.~ (\ref{eq:typical}), essentially operates as a one-variable regression problem, predicting radiance values from different viewpoints. In contrast, our conditional radiance field reconstruction approach incorporates a rotating light environment (in the object frame), making it a regression problem involving two variables. The key insight is to fit a radiance field as a function of both camera views and light rotations, and interpret the NVS problem as a one-dimensional slice of this field.

Either 3DGS or NeRF variants can be used as the baseline 3D representation with our capturing pipeline. In the following, we choose 3DGS for demonstration, and we also provide results for a NeRF-based variant in our supplementary materials. In standard 3DGS, radiance is represented using spherical harmonics (SH) basis functions; making them conditional on varying light rotations is not convenient. Instead, we use a neural network conditioned on light rotations (in addition to view directions). Specifically, each Gaussian has an appearance latent (feature) vector, which can be interpreted by an MLP. The final radiance (color) at each Gaussian is computed as:
\begin{equation}
    c = \mathrm{MLP}(x | v, \theta), 
\end{equation}
where $x$ is the latent vector stored in each Gaussian point, $v$ is the view direction (a normalized 3D direction), and $\theta$ is the light rotation angle. See Sec. \ref{sec:imp} for more details on the MLP architecture.
We evaluate the network before splatting the Gaussians, and the final pixel colors are blended using standard 3DGS differentiable rasterization, which does not require any modifications.

\subsection{Swing sampling for optimal NVS quality}
\label{sec:swing_sampling}

With the \textit{rotating} sampling strategy ($s = 2\pi$), many valid image samples can be obtained. However, when the object rotates, the illumination also rotates, introducing more variation of the radiance at the same time. As a result, the rotation makes the reconstruction of the radiance field more difficult, and not all images from a rotation are equally helpful to the final NVS, since some samples are far from the problem in the 2D sampling domain. Therefore, we need to find an optimal rotating strategy to balance the sample number and the difficulty in reconstruction. In this section, we further investigate the behavior of rotating capturing and find an optimal setting (rotating angle $s$) for our capturing pipeline.

\myfigure{space_swing}{space_swing.pdf}{Our \textit{swing} sampling strategy illustrated in the 2D sampling domain. (a) The object is swinging around the original position on a turntable, and the camera is placed at multiple positions. (b) Different swing angles can lead to different shapes and densities of the sampling region, given the same amount of capturing time.}

To balance between the sample number and NVS quality, the key is to find the most helpful samples, which means to sample near the NVS problem (i.e., the original light condition). Instead of rotating for full circles, we can let the object \textit{swing} around its original position back and forth. In this case, the turntable first rotates for $s$ radians (clockwise) and stops for $m$ seconds. Then, it starts rotating backward (counter-clockwise) for the same angle $s$, stops for another $m$ seconds, and repeats. During each rotation, we still set the camera on the tripod to capture video sequences, and during the $m$ seconds between each rotation, we manually relocate the tripod and camera to get ready for the next round. We also illustrate the \textit{swing} sampling strategy in Fig.~\ref{fig:space_swing} (a).

We can compute the total time cost $t$ of such a \textit{swing} capture as:
\begin{equation}
    T = M\frac{s}{v} + (M - 1) m, s \in [0, 2\pi],
\end{equation}
where $v$ is the angular speed of the turntable. The total number of valid samples can also be represented as:
\begin{equation}
    P = Msn,
\end{equation}
where $n$ is the number of samples per radian. With the formulation above, given the same time cost $t$, we can derive and conclude that:
\begin{itemize}
    \item A larger swing angle generates more samples.
    \item   A larger swing angle involves less human labor, since we need to relocate the camera for $M-1$ times. 
    \item A larger swing angle makes a wider sampling region in the 2D sampling domain, leading to images that are less helpful to the final NVS quality. We also illustrate the effect of different swing angles in Fig.~\ref{fig:space_swing} (b).
    \item When $s=0$, this \textit{swing} strategy degrades to \textit{static} sampling (Sec.~\ref{sec:static_sampling}). 
    \item When $s=2\pi$, this \textit{swing} strategy degrades to \textit{rotating} sampling (Sec.~\ref{sec:rotating_sampling}).
\end{itemize}

Unfortunately, it is nontrivial to find an analytical solution for the optimal swing angle $\hat{s}$. However, with experiments, we show the relationship between swing angle $s$ and the final NVS quality in Fig.~\ref{fig:curve_swing} and the peak signal-to-noise ratio (PSNR) values in Table~\ref{tab:swing_angle}. As a conclusion, we find that $s^\ast=0.2\pi$ achieves a good balance between manual labor and NVS quality, and we also provide some rendering comparisons in our supplementary. 


\begin{table}[h]
    \centering
    \caption{
    The comparison of NVS qualities with different swing angles in the same capturing time (2 minutes). The qualities are evaluated in PSNR ($\uparrow$), and we take the \textsc{Clown} scene as an example. Overall, $s^\ast=0.2\pi$ achieves a good balance between the manual labor, sample number and final NVS quality.
    }
    \label{tab:swing_angle}
    \begin{tabular}{ccc}
    \toprule
         Swing angle $s$&Total images $P$& NVS quality (PSNR)\\
     \midrule
         0 (\textit{static})&50& 37.73\\
 $0.05\pi$& 54&36.34\\
         $0.1\pi$&96& 38.47\\
         $\mathbf{0.2\pi}$&\textbf{140}& \textbf{38.80}\\
         $0.25\pi$&162& 38.08\\
         $0.5\pi$&234& 36.79\\
         $\pi$&324& 36.61\\
         $2\pi$&360& 37.17\\
     \bottomrule
    \end{tabular}
\end{table}

\myfigure{curve_swing}{curve_swing.pdf}{The trends of the manual labor, sample number and sample effectiveness, and their combined effects on the final NVS quality when increasing the swing angle. We observe that $s^\ast = 0.2\pi$ produces the highest NVS quality, achieving a good balance among all these factors.}

\section{Applications}
\label{sec:application}

With our proposed capturing pipeline, we can capture images and predict not only the NVS results at the original light condition, but also any light rotations at any views, allowing for extensive applications in different scenarios. Here, we outline several of these applications, and we also demonstrate them in Sec.~\ref{sec:res_application}.

\paragraph{NVS with light rotations}
One obvious application of our pipeline is for NVS, by predicting the radiance field at any given light rotation. Additionally, our pipeline also naturally supports adjusting the light condition for objects at given views, which can be used in tasks like harmonization. Note that our prediction only works for the rotation angles within the range of captured data. Therefore, in terms of adjusting light conditions, we suggest that users increase the swing angle to have more flexibility.

\paragraph{NVS with distilled SH representation}
Besides using the neural representation directly, it can be distilled into other representations (e.g., SHs) by fixing a light rotation angle. The distilled representation allows novel view synthesis, while being compatible with existing 3DGS-based applications.

\paragraph{Novel lights from combination of rotations}
Since our neural representation encodes the radiance field with multiple light conditions, we can relight the object by combining different light rotations (e.g., front-lit and back-lit ones), achieving extensive light conditions. Furthermore, if the original environment lighting is known, it is possible to approximately fit a target lighting using the known rotations, in order to achieve even more complex relighting tasks.


\section{Implementation details}
\label{sec:imp}





\paragraph{Capturing and dataset setups}

We use the 48 mm camera from an iPhone 15 Pro Max for capturing, and the snapshot of our capturing environment is shown in Fig.~\ref{fig:teaser} (a).
We provide both the \textit{rotating} dataset (i.e., $s=2\pi$) and \textit{swing} dataset (i.e., $s=0.2\pi$) for extensive validation and downstream research. Both dataset consists of 35 real-world objects with diffuse, furry, or glossy appearances. 
For \textit{rotating} capturing, we capture $M=8$ turntable videos and select $N=60$ frames from each video, resulting in a total of 480 images. For \textit{swing} capturing, we set $M=20$ and $N=7$, resulting in 140 images. 
During capturing, we position the camera approximately every $\frac{\pi}{4}$ radians around the object, while varying elevation angles. 
To obtain sufficient samples, a capturing process takes us from 2 minutes ($s=0.2\pi$) to 3 minutes ($s=2\pi$). 

\paragraph{Dataset preparation}

After capturing, we crop the video into $1440\times1440$ rotation segments and randomly select frames from each rotation. Since the turntable rotates at a fixed angular speed, we divide the rotation angle by the number of frames to determine the rotation angle at each selected timestamp. We then conduct calibration using COLMAP \cite{schonberger2016structure}, and finally remove the background and mask them using SAM2 \cite{ravi2024sam}. Compared to the traditional \textit{static} capturing, our pipeline does not introduce any extra procedures in data preparation. Preparing a dataset of an object usually takes us 30 minutes. 

\paragraph{MLP architecture and optimization}
 
In our neural radiance representation, the MLP is 128-channel with 2 hidden layers, using a level-1 frequency encoding for the input view directions and light rotations. The latent vectors are 8-dimensional. During optimization, the network weights and the latent vectors are jointly updated and optimized to fit the appearance under the rotating light conditions. We use the same loss functions from the standard 3DGS.

\section{Results}
\label{sec:results}

\subsection{Experiment setup}

We validate our pipeline on our captured and synthetic datasets. We run all experiments on an Ubuntu 22.04 LTS distribution powered by Windows Subsystem Linux 2. The optimization typically takes about 5-10 minutes on an RTX4090 GPU. 
We implement our neural radiance representation based on 3DGS~\cite{kerbl20233d} for comparison. There are other advanced approaches~\cite{meng2024mirror, liu20243dgs, yu2024gsdf} that can enhance the NVS quality, while they are orthogonal to our pipeline. We did not perform comparisons with them, since it is straightforward to implement our method based on any 3DGS-based NVS approaches.



\subsection{Quality validation}

\begin{table}[htb]
    \centering
    \caption{ 
    The NVS quality of our neural radiance representation and 3DGS on our captured datasets. We choose $s=0.2\pi$ for the best NVS quality. All testing sets are with \textit{static} capturing strategy. The results are evaluated in PSNR~$(\uparrow)$, and the best results are colored in \colorbox{red!20}{red}. 
    }
    \label{tab:ours_vs_3dgs_swing}
    \begin{tabular}{lcclcc}
\toprule
Scene            & Ours   &\multicolumn{1}{l|}{3DGS}    &  Scene           & Ours   & 3DGS  \\
\midrule
Alpaca           & \winner34.48  & 33.11 & \multicolumn{1}{|l}{Jar}              & \winner35.40  & 34.24  \\
Bear             & \winner35.68  & 35.03 & \multicolumn{1}{|l}{JoyCon}           & \winner37.43  & 36.76  \\
BirdGreen        & \winner34.84  & 33.71 & \multicolumn{1}{|l}{Kirby}            & \winner38.20  & 37.54  \\
BirdPink         & \winner35.25  & 34.80 & \multicolumn{1}{|l}{Kodak}            & \winner34.42  & 33.06  \\
Burger           & \winner37.92  & 36.22 & \multicolumn{1}{|l}{Monza}            & \winner38.36  & 37.46  \\
Cap              & \winner28.55  & 26.31 & \multicolumn{1}{|l}{Mug}              & \winner31.08  & 30.22  \\
Capybara         & \winner36.09  & 35.21 & \multicolumn{1}{|l}{Octopus}          & \winner40.35  & 38.70  \\
CapySmall        & \winner38.71  & 38.41 & \multicolumn{1}{|l}{Panda}            & \winner35.19  & 33.06  \\
Cat              & \winner34.22  & 33.73 & \multicolumn{1}{|l}{Pearl}            & \winner38.47  & 37.02  \\
China            & \winner37.71  & 36.96 & \multicolumn{1}{|l}{Penguin}          & \winner34.73  & 33.20  \\
Clown            & \winner38.42  & 35.92 & \multicolumn{1}{|l}{Pine}             & \winner31.63  & 29.69  \\
Controller       & \winner34.83  & 34.00 & \multicolumn{1}{|l}{Rabbit}           & \winner30.69  & 29.83  \\
Dinosaur         & \winner41.80  & 39.66 & \multicolumn{1}{|l}{Rider}            & \winner37.04  & 36.33  \\
Dolphin          & \winner39.35  & 38.16 & \multicolumn{1}{|l}{RiderSmall}       & \winner38.98  & 37.49  \\
Eggplant         & \winner36.09  & 34.30 & \multicolumn{1}{|l}{Sparrow}          & \winner35.13  & 33.83  \\
Ferrari          & \winner36.52  & 35.45 & \multicolumn{1}{|l}{Sunglasses}       & \winner33.91  & 33.71  \\
Fries            & \winner39.38  & 38.69 & \multicolumn{1}{|l}{Truck}            & \winner37.42  & 35.31  \\
HanSolo          & \winner37.45  & 36.29 & \multicolumn{1}{|l}{Tuan}             & \winner36.01  & 35.00  \\
\midrule
\textit{Average} & \winner36.15& 34.95& & & \\
\bottomrule
    \end{tabular}
\end{table}

\begin{table}[htb]
    \centering
    \caption{ 
    The rendering quality at novel light rotations of our neural radiance representation and 3DGS on our captured datasets. We choose $s=2\pi$ to best support light rotations. All testing sets are with \textit{rotating} capturing strategy, where the rotating angles vary from 0 to $2\pi$. The results are evaluated in PSNR~$(\uparrow)$, and the best results are colored in \colorbox{red!20}{red}. 
    }
    \label{tab:ours_vs_3dgs_rotating}
    \begin{tabular}{lcclcc}
\toprule
Scene            & Ours   &\multicolumn{1}{l|}{3DGS}    &  Scene           & Ours   & 3DGS  \\
\midrule
Alpaca           & \winner31.83  & 29.83 & \multicolumn{1}{|l}{Jar}              & \winner27.82  & 23.48  \\
Bear             & \winner33.50  & 25.67 & \multicolumn{1}{|l}{JoyCon}           & \winner37.74  & 30.54  \\
BirdGreen        & \winner32.94  & 23.96 & \multicolumn{1}{|l}{Kirby}            & \winner36.44  & 29.55  \\
BirdPink         & \winner32.90  & 27.05 & \multicolumn{1}{|l}{Kodak}            & \winner30.43  &        27.51\\
Burger           & \winner37.20  & 29.80 & \multicolumn{1}{|l}{Monza}            & \winner36.11  & 29.54  \\
Cap              & \winner27.43  & 24.97 & \multicolumn{1}{|l}{Mug}              & \winner29.71  & 26.56  \\
Capybara         & \winner33.16  & 22.79 & \multicolumn{1}{|l}{Octopus}          & \winner36.61  & 27.54  \\
CapySmall        & \winner34.70  & 25.04 & \multicolumn{1}{|l}{Panda}            & \winner28.58  & 25.80  \\
Cat              & \winner30.35  & 26.01 & \multicolumn{1}{|l}{Pearl}            & \winner37.26  & 29.09  \\
China            & \winner33.95  & 30.07 & \multicolumn{1}{|l}{Penguin}          & \winner28.80  & 24.00  \\
Clown            & \winner39.95  & 35.96 & \multicolumn{1}{|l}{Pine}             & \winner24.16  & 21.86  \\
Controller       & \winner33.83  & 29.73 & \multicolumn{1}{|l}{Rabbit}           & \winner30.05  & 22.99  \\
Dinosaur         & \winner35.25  & 32.52 & \multicolumn{1}{|l}{Rider}            & \winner28.12  & 25.41  \\
Dolphin          & \winner38.03  & 33.51 & \multicolumn{1}{|l}{RiderSmall}       & \winner35.56  & 31.04  \\
Eggplant         & \winner37.82  & 27.96 & \multicolumn{1}{|l}{Sparrow}          & \winner29.72  & 25.53  \\
Ferrari          & \winner32.37  & 30.63 & \multicolumn{1}{|l}{Sunglasses}       & \winner32.20  & 28.75  \\
Fries            & \winner34.96  & 28.77 & \multicolumn{1}{|l}{Truck}            & \winner35.44&        25.62\\
HanSolo          & \winner36.06  & 31.88 & \multicolumn{1}{|l}{Tuan}             & \winner30.79  & 23.86  \\
\midrule
\textit{Average} &        \winner33.10&       27.63& & & \\
\bottomrule
    \end{tabular}
\end{table}

    \begin{table}
    \caption{ The comparison of NVS qualities on synthetic datasets. All results are evaluated in PSNR~$(\uparrow)$, and the top-3 results are marked in \colorbox{red!20}{red}/\colorbox{orange!20}{orange}/\colorbox{yellow!20}{yellow}. Note that in this comparison, the testing set is \textit{static}. Our method produces higher-quality results from training images with rotating lights than 3DGS does with common dataset setups (100 static images), and our \textit{swing} sampling strategy produces the overall highest quality, even compared to 3DGS with fairly sufficient ideally sampled training images.}
    \label{tab:ours_vs_3dgs_static}
        \centering
        \begin{tabular}{l|cccc} 
    \toprule
\multirow{2}{*}{Scene}&  Ours (140&  Ours (480&  3DGS (100& 3DGS (480\\ 
 &  swing)& 

rotating)&  static)& static)\\ 
    \midrule
Armadillo       & \winner36.73 & \normal35.63& \thirdp35.94 & \runner36.52\\
Ficus           & \winner35.45 & \thirdp34.15& \normal34.00 & \runner35.12\\
Flowers         & \winner32.05 & \runner31.89& \normal30.23 & \thirdp31.24\\
Lego            & \winner33.89 & \runner33.32& \normal32.13 & \thirdp32.47\\
    \midrule
Average         & \winner34.53 & \thirdp33.74& \normal33.07 & \runner33.83\\ 
\bottomrule
        \end{tabular}
        \label{tab:my_label}
    \end{table}

The validation of our method includes two aspects: the NVS and the light rotation. Since we can choose different swing angles or different purposes, we provide both the \textit{rotating} dataset ($s=2\pi$) and \textit{swing} dataset ($s=0.2\pi$) for validation. 

For NVS quality, we compare with traditional NVS methods on our \textit{swing} datasets in Table~\ref{tab:ours_vs_3dgs_swing} and Fig.~\ref{fig:ours_vs_3dgs_swing}. The training set is captured with swing angle $s=0.2\pi$ and the testing set is \textit{static}. 
Our method can produce higher quality compared to existing NVS methods on all scenes. 

For the rendering quality of light rotations, we compare with traditional NVS methods on our \textit{rotating} datasets in Table~\ref{tab:ours_vs_3dgs_rotating} and Fig.~\ref{fig:ours_vs_3dgs_rotating}. The training set is captured with rotating angle $s=2\pi$, and the testing set is also \textit{rotating}. 
By comparison, we find that our method has a much higher quality consistently. In particular, 3DGS exhibits an overly dark appearance at grazing angles and obvious blurriness on the furry surface, as it cannot handle rotating light conditions. Although they can reconstruct good geometries, the radiance (especially the shadow effects) is incorrectly predicted. In contrast, our method produces results that closely match the reference, thanks to the conditional radiance representation.




In Fig.~\ref{fig:teaser} (b), we compare the NVS quality from our \textit{swing} capturing with that from the traditional \textit{static} capturing in equal capturing time (2 minutes). We apply our neural radiance representation on both datasets to show the difference in the qualities of captured data. We set the rotation angle to be zero for the traditional capturing. Our capturing pipeline is not only easier to perform, but also producing higher-quality reconstruction with more detailed appearances.

In Table~\ref{tab:ours_vs_3dgs_static} and Fig.~\ref{fig:ours_vs_3dgs_static}, we compare the NVS quality with 3DGS, assuming that we provide it with ideally captured datasets.
Our neural radiance representation achieves higher-quality renderings than those from 3DGS with a common setup (100 static images) and is still competitive with results of 3DGS from abundantly sampled static views (480 static images). 
Note that in practice, even for 480 images with rotating lights, we only need less than 4 minutes to capture them, while capturing 100 high-quality static images already costs more than 4 minutes of difficult manual labor. 



\myfigure{ours_vs_3dgs_static}{ours_vs_3dgs_static.pdf}{Comparison of NVS results by our captured datasets and 3DGS with ideally sampled \textit{static} datasets. The best/second-best results are marked as \textbf{bold}/\textit{italic}. Our neural radiance representation achieves high-quality renderings, and is competitive even with results of 3DGS from abundantly sufficient training images.}



In our supplementary material, we provide ablation studies on two crucial factors (the number of camera positions $M$ and the swing angle $s$). In addition, we also show the optimal swing angles under different frequencies of the environment lights. Please refer to them for more details.

\subsection{Validation on applications}
\label{sec:res_application}


\paragraph{NVS with light rotations}
In Figs.~\ref{fig:teaser} (c) and ~\ref{fig:relight}, we showcase the rendered result under different light rotations. With varying lighting conditions, our rendered results exhibit overall reasonable lighting outcomes. We also show the harmonization results in Fig.~\ref{fig:harmonization} and demonstrate this application in our supplementary video. 

\paragraph{NVS with distilled SH representation}
After selecting a lighting angle, our neural representation can be easily distilled into an SH representation. This way, it can fit existing 3DGS-based applications. 
In Fig.~\ref{fig:distill}, we show the NVS results rendered with the distilled SHs under three different lighting rotations after a 3-minute distillation process. The distilled SHs achieve close qualities to the reference at the specified rotation angles. Note that all the results are obtained from only one training pass. 

\paragraph{Novel light from combination of rotations}
Our neural representation encodes the radiance field with multiple light conditions, which allows us to combine different light rotations (e.g., front-lit and back-lit ones), achieving a fused light condition. In Fig.~\ref{fig:combination}, we demonstrate the rendering results by combining the front-lit and the back-lit scenarios, with different lighting colors. This way, our method enables the creation of rich and diverse assets.

\subsection{Discussion and limitations}

\paragraph{More flexible relighting}
Our capture pipeline enables NVS with different light rotations or a combination of various lighting conditions. However, our method still cannot fully or flexibly support relighting in entirely new environments. We will address this in future work.

\paragraph{Data processing}
Our capture pipeline relies on COLMAP~\cite{schonberger2016structure} to calibrate our camera poses. However, the changing light condition during the capture slightly raises the difficulty of calibration. While using a checkerboard to carry the object can partially alleviate this issue, a COLMAP-free calibration approach will further improve our capture quality. Besides, like traditional capturing methods, the final quality of our datasets might be occasionally hurt due to the inconsistent segmentation from the large model \cite{ravi2024sam}. Therefore, a more accurate background removal approach will also help our pipeline.

\paragraph{Alternative NVS frameworks}
We chose the basic 3DGS framework to validate the effectiveness of our capturing pipeline. However, there are still many advanced approaches that provide even higher NVS qualities. Their work is orthogonal to ours, and it is straightforward to adapt them into our pipeline to achieve advanced NVS qualities.

\section{Conclusion}

In this paper, we have presented \textit{Free Your Hands}, a lightweight object-capturing pipeline to reduce manual workload, standardize the acquisition process, and ensure repeatability. The proposed capture pipeline consists of a simple setup: a consumer turntable to hold the target object and a tripod to hold the camera. As the turntable rotates, we can easily capture hundreds of valid images in several minutes without hands-on effort, minimizing human errors. Then, we design a neural radiance representation conditioned on light rotations tailored for the captured images, as well as an optimal rotation configuration for our pipeline in terms of the final NVS qualities. Our capture pipeline can be integrated into both NeRF- and 3DGS-based frameworks. We have demonstrated the effectiveness of our pipeline in the 3DGS-based framework across various applications, including NVS under different light rotations or combined lighting conditions and harmonization in novel light conditions, showing higher quality in NVS.


There are still many potential future research directions. One promising avenue is to expand the neural radiance representation into a fully relightable representation. Additionally, improving the NVS and relighting quality on some difficult types of objects, such as reflective or transparent ones, is also an interesting and challenging direction.

\clearpage
\bibliographystyle{ACM-Reference-Format}
\bibliography{paper}

\clearpage

\clearpage

\mycfigure{ours_vs_3dgs_swing}{ours_vs_3dgs_swing.pdf}{Comparison of NVS results by our neural radiance representation and 3DGS on our \textit{swing} datasets. The best results are marked as \textbf{bold}. Our capturing pipeline can produce higher-quality NVS results compared to 3DGS, and the whole capturing process is without much manual efforts.}

\mycfigure{ours_vs_3dgs_rotating}{ours_vs_3dgs_rotating.pdf}{Comparison of rendering results with novel light rotations by our neural radiance representation and 3DGS on our \textit{rotating} datasets. The best results are marked as \textbf{bold}. 3DGS cannot handle the rotating light conditions, resulting in wrongly predicted shadow/light effects. In contrast, our pipeline provides closer results to the reference. }

\myfigure{relight}{relight.pdf}{The rendering results of our model from a novel view and novel light rotations. Our method can effectively predict reasonable light transition when light rotates, as our light-conditioned neural representation is learned based on samples from multiple light conditions. }

\myfigure{harmonization}{harmonization.pdf}{Since our conditional neural representation encodes light rotations, we can achieve harmonization by simply adjusting the rotation angle to find the best-suitable appearance of lights and shadows.}

\myfigure{combination}{combination.pdf}{Relighting the objects by linear combinations of light rotations with RGB weights. Top: blue and purple light combined. Bottom: yellow and green light combined.}

\myfigure{distill}{distill.pdf}{The NVS result with SHs distilled from our trained model. By a simple distillation process, our conditional radiance representation can be exported into a static radiance field at a specific light rotation angle for simple NVS and easy cooperation with other applications.}


\clearpage

\end{document}


\title{Supplementary Materials: \\Free Your Hands: Lightweight Turntable-Based Object Capture Pipeline}


\author{Jiahui Fan}
\orcid{0000-0003-0871-7615}
\affiliation{
    \institution{Nanjing University of Science and Technology}
    \country{China}
}
\email{fjh@njust.edu.cn}

\author{Fujun Luan}
\orcid{0000-0001-5926-6266}
\affiliation{
    \institution{Adobe Research}
    \country{USA}
}
\email{fluan@adobe.com}

\author{Jian Yang$^\dagger$}
\orcid{0000-0003-4800-832X}
\affiliation{
    \institution{Nanjing University of Science and Technology}
    \country{China}
}
\email{csjyang@njust.edu.cn}

\author{Milo\v{s} Ha\v{s}an}
\orcid{0000-0003-3808-6092}
\affiliation{
    \institution{Adobe Research}
    \country{USA}
}
\email{milos.hasan@gmail.com}

\author{Beibei Wang$^\dagger$}
\orcid{0000-0001-8943-8364}
\affiliation{
    \institution{Nanjing University}
    \country{China}
}
\email{beibei.wang@nju.edu.cn}

\renewcommand\shortauthors{Fan J. et al}

\maketitle

In the main paper, we have proposed \textit{Free Your Hands}, a lightweight turntable-based object capturing pipeline. This pipeline allows for rendering with novel views and light rotations. We have also introduced a novel formulation of NVS problems in a 2D sampling space, and managed to find an optimal sampling strategy to build our pipeline.
In this supplementary, we provide:
\begin{itemize}
\item  results with a NeRF-based underlying representation \cite{chen2022tensorf},
    \item ablation studies on two crucial factors (number of camera positions $M$ and swing angle $s$) regards the final NVS quality,
    \item and the trends of optimal swing angles with the environment lighting of different frequencies.
\end{itemize}

\section{NeRF-based implementation}

\begin{table}[htb]
    \centering
    \caption{ 
    The NVS quality of our neural radiance representation and TensoRF on our captured datasets. We choose $s=0.2\pi$ for the best NVS quality. All testing sets are with \textit{static} capturing strategy. The results are evaluated in PSNR~$(\uparrow)$, and the best results are colored in \colorbox{red!20}{red}.
    }
    \label{tab:ours_vs_nerf_swing}
    \begin{tabular}{lcc|lcc}
\toprule
Scene & Ours & TensoRF &  Scene & Ours & TensoRF \\
\midrule
Alpaca           & \winner39.05 & 38.58 & \multicolumn{1}{|l}{Jar}              & \winner40.14 & 39.54 \\
Bear             & \winner40.83 & 40.37 & \multicolumn{1}{|l}{JoyCon}           & \winner37.49 & 36.94 \\
BirdGreen        & \winner36.94 & 35.52 & \multicolumn{1}{|l}{Kirby}            & \winner44.32 & 43.88 \\
BirdPink         & \winner41.27 & 40.69 & \multicolumn{1}{|l}{Kodak}            & \winner33.02 & 32.60 \\
Burger           & \winner38.75 & 38.33 & \multicolumn{1}{|l}{Monza}            & \winner43.23 & 42.78 \\
Cap              & \winner38.91 & 38.45 & \multicolumn{1}{|l}{Mug}              & \winner30.96 & 30.43 \\
Capybara         & \winner39.77 & 39.17 & \multicolumn{1}{|l}{Octopus}          & \winner41.67 & 41.08 \\
CapySmall        & \winner30.83 & 30.35 & \multicolumn{1}{|l}{Panda}            & \winner37.36& 37.03 \\
Cat              & \winner36.05 & 35.54 & \multicolumn{1}{|l}{Pearl}            & 40.74& \winner40.77\\
China            & \winner42.80 & 42.32 & \multicolumn{1}{|l}{Penguin}          & \winner36.17 & 35.71 \\
Clown            & \winner43.28&       43.01& \multicolumn{1}{|l}{Pine}             & \winner35.25 & 34.66 \\
Controller       & \winner38.96 & 38.38 & \multicolumn{1}{|l}{Rabbit}           & \winner34.56 & 34.05 \\
Dinosaur         & \winner44.22 & 43.76 & \multicolumn{1}{|l}{Rider}            & \winner37.98 & 37.48 \\
Dolphin          & \winner44.37 & 43.93 & \multicolumn{1}{|l}{RiderSmall}       & \winner39.12 & 38.78 \\
Eggplant         & \winner38.20&       37.98& \multicolumn{1}{|l}{Sparrow}          & \winner36.76 & 36.49 \\
Ferrari          & \winner35.82 & 35.27 & \multicolumn{1}{|l}{Sunglasses}       & \winner32.46 & 32.08 \\
Fries            & \winner41.64 & 41.26 & \multicolumn{1}{|l}{Truck}            & \winner39.07 & 38.45 \\
HanSolo          & \winner35.14 & 34.54 & \multicolumn{1}{|l}{Tuan}             & \winner39.51 & 38.94 \\
\midrule
\textit{Average} & \winner38.51& 38.03& & & \\
\bottomrule
    \end{tabular}
\end{table}

\begin{table}[htb]
    \centering
    \caption{ 
    The rendering quality at novel light rotations of our neural radiance representation and TensoRF on our captured datasets. We choose $s=2\pi$ to best support light rotations. All testing sets are with \textit{rotating} capturing strategy, where the rotating angles vary from 0 to $2\pi$. The results are evaluated in PSNR~$(\uparrow)$, and the best results are colored in \colorbox{red!20}{red}.
    }
    \label{tab:ours_vs_nerf_rotating}
    \begin{tabular}{lcc|lcc}
\toprule
Scene & Ours & TensoRF &  Scene & Ours & TensoRF \\
\midrule
Alpaca           & \winner36.24 & 26.10 & \multicolumn{1}{|l}{Jar}              & \winner30.40 & 25.05 \\
Bear             & \winner36.70 & 26.89 & \multicolumn{1}{|l}{JoyCon}           & \winner36.34 & 30.99 \\
BirdGreen        & \winner35.77 & 28.52 & \multicolumn{1}{|l}{Kirby}            & \winner41.29 & 32.77 \\
BirdPink         & \winner37.05 & 26.44 & \multicolumn{1}{|l}{Kodak}            & \winner29.40 & 23.73 \\
Burger           & \winner39.45 & 33.06 & \multicolumn{1}{|l}{Monza}            & \winner38.44 & 27.94 \\
Cap              & \winner32.38 & 16.57 & \multicolumn{1}{|l}{Mug}              & \winner29.42 & 25.85 \\
Capybara         & \winner34.90 & 24.39 & \multicolumn{1}{|l}{Octopus}          & \winner39.88 & 30.37 \\
CapySmall        & \winner28.66 & 25.32 & \multicolumn{1}{|l}{Panda}            & \winner28.02 & 25.82 \\
Cat              & \winner31.54 & 23.88 & \multicolumn{1}{|l}{Pearl}            & \winner38.17 & 27.25 \\
China            & \winner34.91 & 28.57 & \multicolumn{1}{|l}{Penguin}          & \winner28.35 & 22.36 \\
Clown            & \winner40.59 & 34.47 & \multicolumn{1}{|l}{Pine}             & \winner27.65 & 20.11 \\
Controller       & \winner36.91 & 20.69 & \multicolumn{1}{|l}{Rabbit}           & \winner31.72 & 22.50 \\
Dinosaur         & \winner34.19 & 30.89 & \multicolumn{1}{|l}{Rider}            & \winner28.58 & 24.97 \\
Dolphin          & \winner37.81 & 33.92 & \multicolumn{1}{|l}{RiderSmall}       & \winner37.09 & 26.71 \\
Eggplant         & \winner39.00 & 28.92 & \multicolumn{1}{|l}{Sparrow}          & \winner29.74 & 21.32 \\
Ferrari          & \winner31.22 & 30.11 & \multicolumn{1}{|l}{Sunglasses}       & \winner32.18 & 28.78 \\
Fries            & \winner37.28 & 29.06 & \multicolumn{1}{|l}{Truck}            & \winner35.89 & 26.15 \\
HanSolo          & \winner34.28 & 31.32 & \multicolumn{1}{|l}{Tuan}             & \winner30.84 & 26.61 \\
\midrule
\textit{Average} & \winner34.23 & 26.90 & & & \\
\bottomrule
    \end{tabular}
\end{table}


\myfigure{ablation_M}{ablation_M.pdf}{The effect of the number of camera positions ($M$) in our pipeline. Given only few camera position, we observe one ``diagonal line'' in the 2D sampling space, resulting in severe over-fitting. In practice, we choose $M=8$ for our \textit{rotating} datasets and $M=20$ for \textit{swing} datasets.}
\myfigure{ablation_s}{ablation_s.pdf}{The NVS results of different swing angle with the same capturing time (2 minutes). By comparison, $s^\ast = 0.2\pi$ achieves a good balance between the manual labor and final quality.}

Our pipeline is flexible and compatible to both 3DGS- and NeRF-based NVS approaches. We use neural features to represent the radiance, together with an MLP that is able to decode the feature vectors into RGB colors, shared by all neural features. In NeRF, the radiance is already represented by neural features. Therefore, we simply add the light rotation as a conditional input to the MLP decoder. For efficiency, we choose TensoRF \cite{chen2022tensorf} as the baseline method to validate our pipeline, since its training time is much shorter than the original NeRF \cite{mildenhall2021nerf}.

We provide the quantitative results in Tables~\ref{tab:ours_vs_nerf_swing} (for NVS qualities) and \ref{tab:ours_vs_nerf_rotating} (for light rotations). Overall, our method outperforms TensoRF on both datasets.

\section{Ablation study}

There are two key factors that decide the final NVS quality reconstructed from our captured dataset, and we provide ablation studies on them.

In Fig.~\ref{fig:ablation_M}, we show the rendering results of different numbers of camera positions ($M$) with our \textit{swing} and \textit{rotating} captures. For fairness, we provide the same total number of samples by increasing the frame numbers per rotation ($N$). By comparison, increasing the number of camera locations improves the rendering quality. The main reason is that sparse locations lead to uneven sampling on the 2D sampling space, which can lead to overfitting during optimization and hurt the rendering quality.

In Fig.~\ref{fig:ablation_s}, we compare the rendering quality of different swing angles ($s$) within the same capturing time (2 minutes). By comparison, $s^\ast = 0.2\pi$ provides the best rendering quality, as we described in the main text of this paper.

\section{Environment light frequencies}

By introducing the swing angles, we balance between the NVS difficulty and the number of samples. Generally, with a larger swing angle, the radiance variation when rotating objects is more obvious. We experiment and validate our choice of the optimal swing angle with a common indoor environment lighting. However, if the environment light is of low frequency, which means the illumination varies little with rotation, it might leads to different optimal swing angles.

Theoretically, with lower-frequent (i.e., softer) lights, the swing angle should be increased. We validate this by gradually applying a Gaussian blur on an environment map and find their corresponding optimal swing angles. The blurred environment maps and their quantitative results are shown in Fig.~\ref{fig:swing_angle_with_blur}. In conclusion, for common environments, $s\ast = 0.2\pi$ is a suggested choice for balancing the human labor and NVS quality.

\myfigureh{swing_angle_with_blur}{swing_angle_with_blur.pdf}{Relationship between the NVS quality and swing angle in capturing with environment lights of different frequency. When the light becomes softer, the optimal swing angle is also relatively larger (from $0.2\pi$ to $0.5\pi$). Overall, for common environments, $s^\ast = 0.2\pi$ can achieve the best quality.}

\bibliographystyle{ACM-Reference-Format}
\bibliography{paper}